\documentclass[journal]{IEEEtran}
\usepackage{epsfig,amsmath,amssymb,epsf,cite,subfigure,scalefnt}
\usepackage{algorithm,algorithmic}
\usepackage[retainorgcmds]{IEEEtrantools}
\usepackage{graphicx,multicol}
\usepackage{color,soul} 
\usepackage{amsthm}

\newcount{\myi}
\newcommand{\Repeat}[2]{%
    \myi=0
    \loop
        \ifnum\myi<#2
        #1
        \advance\myi by 1
    \repeat
}

\def\cH{{\mathcal{H}}}

\def\cM{{\mathcal{M}}}

\def\cV{{\mathcal{V}}}
\def\cW{{\mathcal{W}}}

\def\ba{{\mathbf{a}}}

\def\bff{{\mathbf{f}}}

\def\bh{{\mathbf{h}}}

\def\bn{{\mathbf{n}}}

\def\bw{{\mathbf{w}}}
\def\bx{{\mathbf{x}}}
\def\by{{\mathbf{y}}}

\def\bW{{\mathbf{W}}}

\def\argmin{\mathop{\mathrm{argmin}}}
\def\argmax{\mathop{\mathrm{argmax}}}

\def\d4{\!\!\!\!}

\def\-{\! - \!}
\def\+{\! + \!}
\def\={\! = \!}
\def\>{\! > \!}

\def\kth{{\ensuremath{k^{\text{th}}}}}

\def\sin{{\text{sin}}}

\def\kplus1{{\ensuremath{(k+1)^{\text{th}}}}}

\def\bhhat{{\ensuremath{\widehat{\bh}}}}

\newtheorem{lemma}{Lemma}

\newcommand{\bet}{\begin{table}}
\newcommand{\eet}{\end{table}}
\newcommand{\btt}{\begin{tabular}}
\newcommand{\ett}{\end{tabular}}
\newcommand{\bec}{\begin{center}}
\newcommand{\eec}{\end{center}}
\newcommand{\bef}{\begin{figure}}
\newcommand{\eef}{\end{figure}}
\newcommand{\beq}{\begin{eqnarray}}
\newcommand{\eeq}{\end{eqnarray}}
\newcommand{\bea}{\begin{array}}
\newcommand{\eea}{\end{array}}

\providecommand{\abs}[1]{\lvert#1\rvert}
\providecommand{\norm}[1]{\lVert#1\rVert}

\def\bhhat{{\ensuremath{\widehat{\bh}}}}
\def\Pmis{{\ensuremath{P^{\text{mis}}_{k+1 \vert k}(\bw_{k+1})}}}

\begin{document}
%
\title{Closed-Loop Beam Alignment for Massive MIMO Channel
Estimation}
%
%
%
\author{Andrew~J.~Duly, \IEEEmembership{Member, IEEE},
Taejoon~Kim, \IEEEmembership{Member, IEEE},
David~J.~Love, \IEEEmembership{Senior~Member, IEEE}, and
James~V.~Krogmeier \IEEEmembership{Member, IEEE}
\thanks{A. J. Duly, D. J. Love, and J. V. Krogmeier are with the School of
Electrical and Computer Engineering, Purdue University, West Lafayette, IN
47907 USA (e-mail: andrew.j.duly@ieee.org; djlove@ecn.purdue.edu;
jvk@ecn.purdue.edu).}
\thanks{T. Kim is with the Department of Electronic Engineering, City University
of Hong Kong, Kowloon, Hong Kong (e-mail: taejokim@cityu.edu.hk).}
\thanks{The work in this paper was partially supported by a grant from City University of Hong Kong (No. 7200347).}
}

\maketitle

\begin{abstract}
	Training sequences are designed to probe wireless channels in order to
	obtain channel state information for block-fading channels. Optimal
	training sounds the channel using orthogonal beamforming vectors
	to find an estimate that optimizes some cost function, such as 
	mean square error. As the number of transmit antennas increases, however, the
	training overhead becomes significant. This creates a need for alternative
	channel estimation schemes for increasingly large transmit arrays. In this
	work, we relax the orthogonal restriction on sounding vectors. The use of a
	feedback channel after each forward channel use during training enables
	closed-loop sounding vector design. A misalignment cost function is
	introduced, which provides a metric to sequentially design sounding vectors. 
	In turn, the structure of the sounding vectors aligns the transmit beamformer
	with the true channel direction, thereby increasing beamforming gain. This
	beam alignment scheme for massive MIMO is shown to improve beamforming gain
	over conventional orthogonal training for a MISO channel.
\end{abstract}


\section{Introduction}
In wireless communications, accurate channel state information plays a central
role in realizing the gains afforded by coherent communications. In light of
this, much literature exists to design training sequences to estimate the
channel. For a block-fading channel model, \cite{Hassibi2003} addressed capacity
maximizing values for various parameters, including the length of the
training phase and the power of the training phase. The error
covariance of the minimum mean square error (MMSE) channel estimate is minimized
when orthogonal waveforms are transmitted from each element.  For orthogonal
transmission from each element, the discrete length of the training signal must
be at least equal to the number of transmit antennas. For massive MIMO systems,
this would lead to training consuming a large fraction of the coherence interval.

Channel estimation with training sequences of length less than the number of
transmit antennas was considered for distributed transmit beamforming systems in
\cite{Zhang2013}. The set of training sequences which minimizes the mean square
error of the MMSE channel estimate was shown to be similar to those maximizing the
expected beamforming gain, which gives a training signal matrix
with orthogonal columns. The optimal full-rank set of receive beamforming
vectors for angle of arrival estimation was derived in \cite{Fuhrmann2008}.  The
channel gain term was treated as a deterministic unknown, and beamforming
vectors were derived to minimize the variance on the angle of arrival estimate.

The literature referenced thus far in this work pertains to a class of open loop
schemes in which the training sequence is predetermined.  These training
sequences cannot be adapted to knowledge gathered of the instantaneous channel.
Alternatively, adaptive training schemes use a feedback link from the receiver
to the transmitter to guide the design of the training sequence. Closed-loop
training across multiple blocks for massive MIMO beamforming systems is
considered in \cite{Love2013}. Assuming a temporally correlated channel, their
algorithm selected the entire training signal for a given block based on channel estimates from previous
blocks. Improvements in average receive SNR were shown by utilizing the
training sequences received in previous channel blocks. Line-of-sight channel
estimation for large arrays in backhaul cellular networks is presented in
\cite{Hur2013}.  Adaptive subspace sampling, where samples from previous
channel uses aid in beamformer design, gave improved beamforming gain over
non-adaptive techniques. 

In this work, we focus on channel estimation for single-user massive MIMO. These
techniques can be extended to multiuser scenarios, a topic of future work. As
the number of antennas grows large, optimal channel estimation no longer remains practical.
Some of the general properties for massive MIMO systems were shown to hold
for arrays between $100$ and $1,000$ antennas in \cite{ScalingUpMIMO13}. We target
practical massive MIMO systems and develop a beam alignment scheme to estimate
MISO channels on the lower end of that range. We consider beamforming for a MISO
channel in a single coherent channel block. A feedback
channel allows adaptation of the training sequence after each channel use. In
general, current systems do not feed back information after every channel use. 
The increased complexity, however, may become necessary for massive MIMO
systems to shorten the training phase. The training sequence is comprised of
sounding vectors, which are sequentially designed in a manner that aligns the estimated channel with the
true instantaneous channel direction, treating the channel gain term as a
nuisance parameter.  This work focuses on frequency division duplexing (FDD)
systems, where channel reciprocity cannot be used for transmit beamforming.
Nonetheless, this scheme remains applicable for time division duplexing (TDD)
systems with no uplink sounding.

%
%
\section{System Setup}\label{sec:system_setup}
Consider a multiple-input single-output (MISO) wireless channel with $M$
transmit antennas and a single receive antenna. In coherent communications, the
wireless channel is considered known and knowledge of the channel at the
receiver aids symbol detection and demodulation. For a given channel use, the
input-output relationship is
\begin{eqnarray*}
	r = \left( \sqrt{\rho} \bx \right)^* \left( \alpha \bh \right) + n
\end{eqnarray*}
where $\bx \in \mathbb{C}^{M \times 1}$ is the transmitted signal with $E\left[
\norm{\bx}^2 \right] = 1$, $\rho$ denotes the transmit power, $r \in \mathbb{C}$
is the received signal, $n \sim
\mathcal{CN}(0,\sigma^2_n)$ the additive noise, $(\cdot)^*$ denotes the
Hermitian transpose, and $\alpha \bh$ is the MISO wireless channel, comprised of
the complex scalar channel gain $\alpha$ and the channel direction vector $\bh$
with $\norm{\bh} = 1$. The channel direction vector $\bh$ describes direction
only and lacks any sort of gain or attenuation.

Transmit beamforming is commonly used in point-to-point MISO systems, but it is
also expected to be critical to multiuser massive MIMO systems. A beamformed
transmit signal is defined as $\bx = \bff s$, where the beamforming vector $\bff$ and complex
symbol $s$ are constrained to $\norm{\bff} = 1$ and $E[\abs{s}^2] = 1$. The
input-output relationship then simplifies to $ r =  \sqrt{\rho} \alpha (\bff^*
\bh) s+ n$, where $\abs{\bff^* \bh}^2$ denotes the beamforming gain.  The
optimal beamformer $\bff_{\text{opt}}$ that maximizes the beamforming gain and
achieves $\abs{\bff_{\text{opt}}^* \bh}^2 = 1$ is $\bff_{\text{opt}} = \bh$.
However, the true channel direction $\bh$ is unknown and must be estimated.

To estimate the wireless channel, assume $K$ channel uses are available for this
task. If the transmit signal for data transmission is restricted to beamforming,
it is natural to restrict the training signal in the same fashion. Let
the sounding vector $\bw_k$ represent the training signal for the $\kth$ channel
use. The receive vector for the first $k$ channel uses of the training phase is
given by
\begin{eqnarray}\label{eq:Rx_vector}
	\by_k = \alpha \sqrt{\rho} \bW_k^* \bh + \bn_k
\end{eqnarray}
where $\bW_k = [\bw_1, \bw_2, \ldots, \bw_k]$ is the set of $k$ sounding
vectors and $\by_k$ is the $k \times 1$ complex receive sample vector.

\begin{figure}[!htbp]
	\centering
 	\includegraphics[width=\columnwidth]{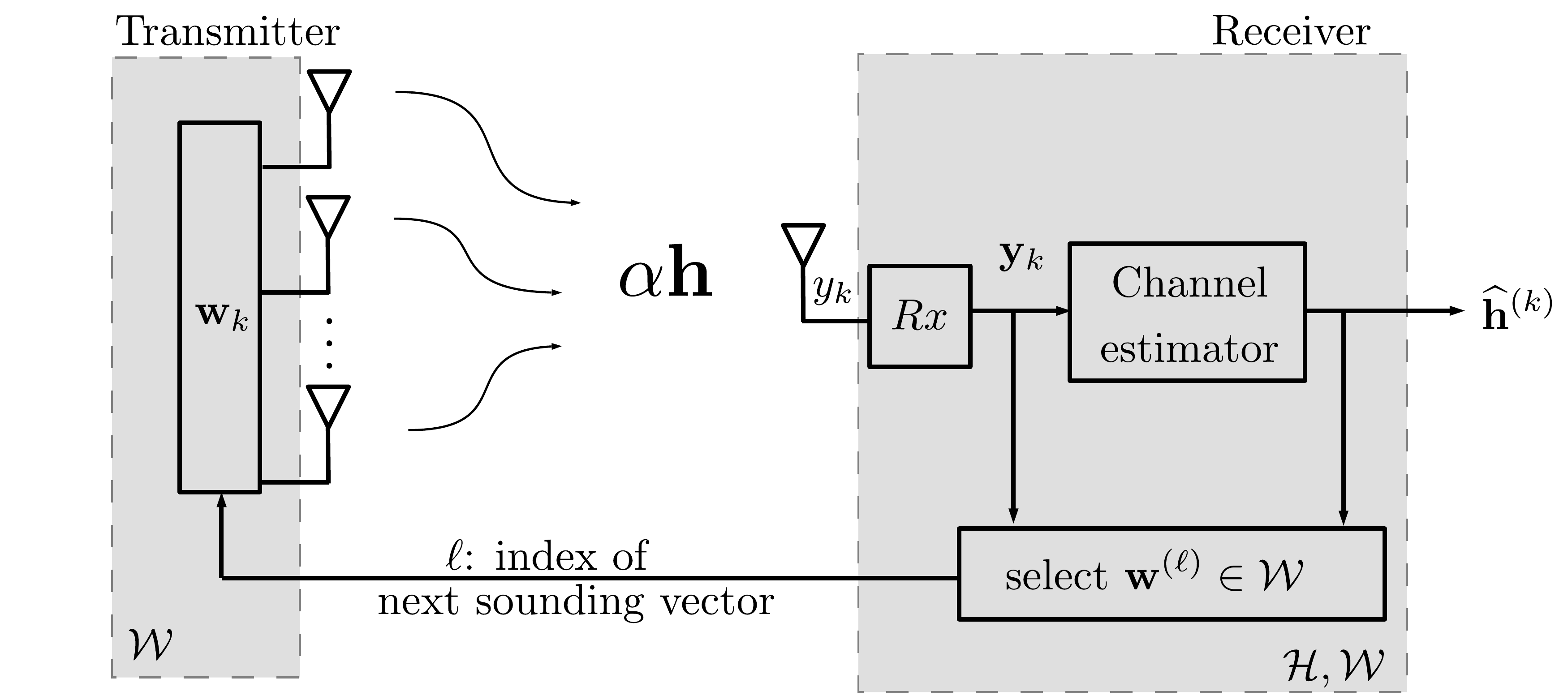}
	\caption{Block diagram illustrating the proposed training scheme. The
	receiver selects the sounding vector from the codebook $\mathcal{W}$ for the
	next channel use and feeds back the codeword index to the transmitter.}
	\label{fig:feedback_bd}
\end{figure}

To aid in channel estimation, we consider a limited-rate feedback channel
from the receiver to the transmitter as shown in Fig. \ref{fig:feedback_bd}. 
Let $\mathcal{W}$ represent the set of possible sounding vectors, $\mathcal{W} =
\{ \bw^{(\ell)} \in \mathbb{C}^{M \times 1}:
\norm{\bw^{(\ell)}} = 1, \ell=1,\ldots,L \}$.
Due to the restricted bandwidth of the feedback channel, we design $\mathcal{W}$ to be a discrete set with $\abs{\mathcal{W}}
= L$, where $\abs{\mathcal{W}}$ denotes the cardinality of the set
$\mathcal{W}$, and make this codebook available ahead of time to both the
transmitter and receiver. For the $\kth$ channel use, the receiver calculates
an estimate of $\bh$ and decides on the appropriate $\bw^{(\ell)} \in
\mathcal{W}$ to sound the $\kplus1$ channel use.  The index of this sounding
vector is fed back to the transmitter, as described in Fig.
\ref{fig:feedback_bd}. Note the feedback link is utilized after every channel
use during training, which differs from previous channel estimation schemes,
including \cite{Love2013}. This creates considerable feedback overhead per
coherence time, which is the tradeoff for fewer pilots.

Assume the channel
direction vector belongs to a discrete set $\cH$.
In practice, $\cH$ approximates the true channel subspace. The maximum
likelihood channel estimator is given as,
\begin{eqnarray*}
	\widehat{\bh}^{(k)}
	&=&\argmax_{\bh \in \mathcal{H}}
	p\left(\by_k \vert \bh, \bW_k, \alpha \right)\\
	&=& \argmin_{\bh \in \mathcal{H}} \norm{\by_k -
	\alpha \sqrt{\rho} \bW_k^* \bh}^2
\end{eqnarray*}
Given this channel direction estimator, after $K$ channel uses of training the
beamforming vector for data transmission is set to $\bff = \widehat{\bh}^{(K)}$.

In addition to estimating the channel direction, the channel gain term $\alpha$
must also be estimated. We treat $\alpha$ as a nuisance parameter and resort
to composite estimation techniques. A generalized maximum likelihood estimator
replaces the true channel gain by its maximum likelihood estimate,
\begin{eqnarray*}
	\widehat{\alpha}_{\bh} &=& \argmax_{\alpha \in \mathbb{C}}
	p\left(\by_k \vert \bh, \bW_k, \alpha \right)
	= \frac{\bh^* \bW_k \by_k}{\sqrt{\rho}\norm{\bW^*_k \bh}^2}.
\end{eqnarray*}
This result can be shown by minimizing the squared norm
$\norm{\by_k - \alpha \sqrt{\rho} \bW_k^* \bh}^2$ over all $\alpha \in
\mathbb{C}$. We write the channel magnitude as a complex scalar to incorporate
the phase invariance of the channel direction estimate. Beamforming gain is
invariant to the phase of the channel direction, e.g., $e^{j \phi}
\widehat{\bh}^{(k)}$ and $\widehat{\bh}^{(k)}$ give the same beamforming gain.
If $\cH$ is a discrete set, a given $\bh \in \cH$ could provide large
beamforming gain but give a low likelihood due to its phase.  This issue is
resolved if we consider $\alpha$ to be complex. The generalized ML channel
estimator is given by substituting $\widehat{\alpha}_{\bh}$ into $p(\by_k \vert 	\bh, \bW_k, \alpha)$,
\begin{eqnarray}\label{eq:cont_ch_est}
	\widehat{\bh}^{(k)} 
	&=& \argmax_{\bh \in \mathcal{H}} \frac{\abs{\by_k^* \bW_k^*
	\bh}^2}{\norm{\bW_k^* \bh}^2} = \argmax_{\bh \in \mathcal{H}} d_k(\bh)
\end{eqnarray}
where $d_k(\bh) = \frac{\abs{\by_k^* \bW_k^* \bh}^2}{\norm{\bW_k^* \bh}^2}$.
In essence, this channel estimator finds the channel vector $\bh \in \mathcal{H}$
whose projection onto $\bW_k$ most closely aligns with the \emph{direction} of
$\by_k$. The normalizing term $\norm{\bW_k^* \bh}^2$ removes any dependence on
vector magnitude.  If one interprets each $\bh \in \cH$ as a beam, the channel
estimator selects the beam whose projection onto $\bW_k$ most closely aligns
with $\by_k$. Beam alignment is performed by properly selecting successive
sounding vectors.
%
%
\section{Sounding Vector Selection}\label{sec:codebook_design}
As one might expect, the sounding vectors have direct influence on the
performance of the channel estimator in \eqref{eq:cont_ch_est}. In this section,
we address how to select the sounding vector for each channel use.
With feedback after every channel use, the sounding vector for the $\kplus1$
channel use can be chosen as a function of the previous $k$ receive samples,
$\by_k$. Let us define the misalignment event after $k+1$ channel uses
as $\cM_i = \big\{ \widehat{\bh}^{(k+1)} \neq \bh  \big\vert \by_k,
\bW_{k+1}, \bh = \bh_i, \widehat{\alpha}_{\bh_i} \big\}$. This misalignment
event occurs when the estimated channel direction $\widehat{\bh}^{(k+1)}$ is not
the true channel direction $\bh = \bh_i$. Averaging the probability of the
misalignment event over all channel directions gives the misalignment cost
function for the $\kplus1$ channel use evaluated at channel use $k$,
\begin{IEEEeqnarray}{rCl}
	\IEEEeqnarraymulticol{3}{l}{
	\Pmis = 
	 \sum_{i=1}^N \text{Pr} (\cM_i) p_i^{(k)}
	 }\IEEEyesnumber\label{eq:pma_general}\\
	\qquad &=& \sum_{i=1}^N \text{Pr} \bigg( d_{k+1}(\bh_i) < \text{max}
	\Big[ \big\{ d_{k+1}(\bh_j) \big\}_{j \neq i} \Big] \Big\vert \cV_i \bigg) p_i^{(k)}
	\nonumber
\end{IEEEeqnarray}
where $p_i^{(k)} = p(\bh_i \vert \by_k, \bW_k, \widehat{\alpha}_{\bh_i})$ are
the updated priors after $k$ channel uses and $\cV_i = \{ \bh = \bh_i,	\by_k, \bW_k,
\widehat{\alpha}_{\bh_i} \}$. This cost function measures the
penalty of choosing the wrong channel direction as a function
of $\bw_{k+1}$. The previous $k$ receive samples are known, and the receive
sample for the $\kplus1$ channel use, $y_{k+1}$, is treated as a
random variable as described in Section \ref{sec:binary_codebook}. In what follows, the misalignment cost
function is adopted to develop a sounding vector selection criterion.

\subsection{Binary Channel Codebook}\label{sec:binary_codebook}
We approximate, in this subsection, the channel using a binary channel codebook,
$\cH = \{ \bh_1, \bh_2 \}$. Despite being an extremely coarse discretization of the channel
space, the exact misalignment cost function can be derived for the binary codebook
case,  
\begin{IEEEeqnarray}{rCl}
	\Pmis
	&=&	\text{Pr} \Big( d_{k+1}(\bh_1) < d_{k+1}(\bh_2) \Big\vert \cV_1
	\Big) p_1^{(k)} \IEEEyesnumber\label{eq:Pm_binary}\\
	&& + \,\, \text{Pr} \Big( d_{k+1}(\bh_2) < d_{k+1}(\bh_1) \Big\vert
	\cV_2 \Big) p_2^{(k)} \IEEEnonumber
\end{IEEEeqnarray}
which is the weighted sum
of pairwise error probability (PEP) terms. The $\text{PEP}\left( \bh_2
\rightarrow \bh_1 \right \vert \cV_2)$ is the probability $\bh_1$ is chosen assuming
$\bh_2$ is true,
\begin{IEEEeqnarray}{lr}
\text{PEP}\left( \bh_2 \rightarrow \bh_1 \right \vert \cV_2) 
= \IEEEyesnumber\label{eq:PEP}\\
\quad\quad \text{Pr} \left( \frac{\abs{\by_{k+1}^* \bW_{k+1}^*
	\bh_1}^2}{\norm{\bW_{k+1}^* \bh_1}^2}  >  \frac{\abs{\by_{k+1}^*
	\bW_{k+1}^* \bh_2}^2}{\norm{\bW_{k+1}^* \bh_2}^2} \Bigg\vert \cV_2  \right).    
\IEEEnonumber
\end{IEEEeqnarray}
Since $\frac{\by_{k+1}^* \bW_{k+1}^* \bh_i}{\norm{\bW_{k+1}^* \bh_i}}$ is a
Gaussian random variable when conditioned on $\bh_i$, the pairwise error
probability is the probability the magnitude of one Gaussian random variable exceeds the magnitude of another. We
now calculate the PEP considering $d_{k+1}(\bh_1)$ and $d_{k+1}(\bh_2)$ are
functions of the same scalar noise term, $n_{k+1}$.
\begin{lemma}\label{lem:PEP}
	Consider two  Gaussian random variables,
	\begin{eqnarray*}
		X &=& \mu_x + q_x n \sim \mathcal{CN}(\mu_x,\abs{q_x}^2)\\
		Y &=& \mu_y + q_y n \sim \mathcal{CN}(\mu_y,\abs{q_y}^2)
	\end{eqnarray*}
	 where $\mu_x,\mu_y,q_x,q_y$ are all constant complex scalars and $n \sim
	\mathcal{CN}(0,1)$. Then,
	\begin{IEEEeqnarray}{rl}
		\IEEEeqnarraymulticol{2}{l}{\text{Pr}\left( \abs{X}^2 > \abs{Y}^2 \right) =
		} \label{eq:P_xy}\\
		\qquad \begin{cases}
			1 -Q_1\left(
	\left\vert \frac{\mu_y q_y^* - \mu_x q_x^*}{\abs{q_y}^2-\abs{q_x}^2}
	\right\vert, \left\vert \frac{\mu_x q_y - q_x \mu_y}{\abs{q_y}^2-\abs{q_x}^2}
	\right\vert \right) & \text{ \emph{if} }
			\abs{q_y}^2 > \abs{q_x}^2 \nonumber\\
			Q_1\left(
	\left\vert \frac{\mu_y q_y^* - \mu_x q_x^*}{\abs{q_y}^2-\abs{q_x}^2}
	\right\vert, \left\vert \frac{\mu_x q_y - q_x \mu_y}{\abs{q_y}^2-\abs{q_x}^2}
	\right\vert \right) & \text{ if }
			\abs{q_y}^2 < \abs{q_x}^2 \nonumber\\
			\frac{1}{2} \left[ 1 + \text{\emph{erf}} \left(
			\frac{\abs{\mu_y}^2-\abs{\mu_x}^2}{2\sqrt{2}\abs{\mu_y^* q_y - \mu_x^* q_x}}
			\right) \right] & \text{ if } \abs{q_y}^2 = \abs{q_x}^2
		\end{cases}
	\end{IEEEeqnarray}
\end{lemma}
where $Q_1(\cdot,\cdot)$ is the first-order Marcum Q function \cite{Proakis}.
\begin{IEEEproof}
	See Appendix.
\end{IEEEproof}
Assuming $\bh_2$ is true, the pairwise error probability in \eqref{eq:PEP} is
calculated using Lemma \ref{lem:PEP} by setting $\mu_x = \frac{\by_k^* \bW_k^* \bh_1 + \sqrt{\rho}
\widehat{\alpha}_{\bh_2}^* \bh_2^* \bw_{k+1} \bw_{k+1}^*
\bh_1}{\norm{\bW_{k+1}^* \bh_1}}$, $q_x = \frac{\bw_{k+1}^*
\bh_1}{\norm{\bW_{k+1}^* \bh_1}}$, $\mu_y = \frac{\by_k^* \bW_k^* \bh_2 +
\sqrt{\rho} \widehat{\alpha}_{\bh_2}^* \abs{\bw_{k+1}^*
\bh_2}^2}{\norm{\bW_{k+1}^* \bh_2}}$, and $q_y = \frac{\bw_{k+1}^*
\bh_2}{\norm{\bW_{k+1}^* \bh_2}}$.

The sounding vector for the next channel use is chosen as,
\begin{eqnarray}\label{eq:sel_soundingvector}
	\bw_{k+1} = \argmin_{\bw^{(\ell)} \in \mathcal{W}}
	P^{\text{mis}}_{k+1 \vert k}(\bw^{(\ell)})
\end{eqnarray}
\subsection{N-ary Channel Codebook}
We now extend our scope to the $N$-ary scenario. Let $\mathcal{H} =
\{\bh_1, \, \bh_2, \, \ldots, \, \bh_N \}$. The misalignment cost function
expression in \eqref{eq:pma_general} cannot be simplified in terms of the
pairwise error probabilities, as was the case for the binary channel codebook. 
Using a union bound argument, we can place an upper bound on the misalignment
cost function
\begin{eqnarray}\label{eq:Pmis_ub}
	\Pmis
	\leq \sum_{i=1}^N \sum_{\substack{j=1\\ j \neq i}}^N \text{PEP}\left(
	\bh_i \rightarrow \bh_j \vert \cV_i \right) p_i^{(k)}. 
\end{eqnarray}
Calculating the misalignment cost function requires evaluating $\abs{\cH}
(\abs{\cH}-1)$ PEP terms for each sounding vector codeword in $\cW$. For larger
codebooks, the complexity of this algorithm becomes significant, as
$\abs{\cH} (\abs{\cH}-1) \abs{\cW}$ PEP terms must be calculated for each
channel use.

\subsection{Low Complexity Beam Alignment}
In general, the size of the channel codebook and sounding vector codebook scales
with the number of transmit antennas to maintain a fixed quantization
error. As a result, the complexity to select a sounding vector at each channel
use will scale with $M$. To make the beam
alignment algorithm tractable for moderately sized arrays, we approximate
\eqref{eq:Pmis_ub}. To do this, we note that
as the number of channel uses increases, the beam alignment algorithm typically prefers a single
codeword corresponding to a high $p_i^{(k)}$. Based on this observation, we can
simplify the misalignment cost function to only consider the two most likely
codewords. Letting $\bh_{(1)}$ represent the most likely channel codeword and $\bh_{(2)}$ represent the second most likely channel codeword, we can approximate \eqref{eq:Pmis_ub} as
\begin{IEEEeqnarray*}{rCl}
	\Pmis &\approx& 
	\text{PEP}\left( \bh_{(1)} \rightarrow \bh_{(2)} \vert \cV_{(1)}
	\right) p_{(1)}^{(k)} \\ 
	&& \quad + \,\, \text{PEP}\left( \bh_{(2)} \rightarrow \bh_{(1)} \vert
	\cV_{(2)} \right) p_{(2)}^{(k)}.
\end{IEEEeqnarray*}
In this manner, the beam alignment algorithm calculates $2 \abs{\cW}$ PEP terms,
a significant reduction from the union bound which
calculates $\abs{\cH} (\abs{\cH}-1) \abs{\cW}$ PEP terms.
%
%
\section{Simulations}\label{sec:simulations}
\begin{figure}[!htbp]
\centering
\includegraphics[width=\columnwidth]{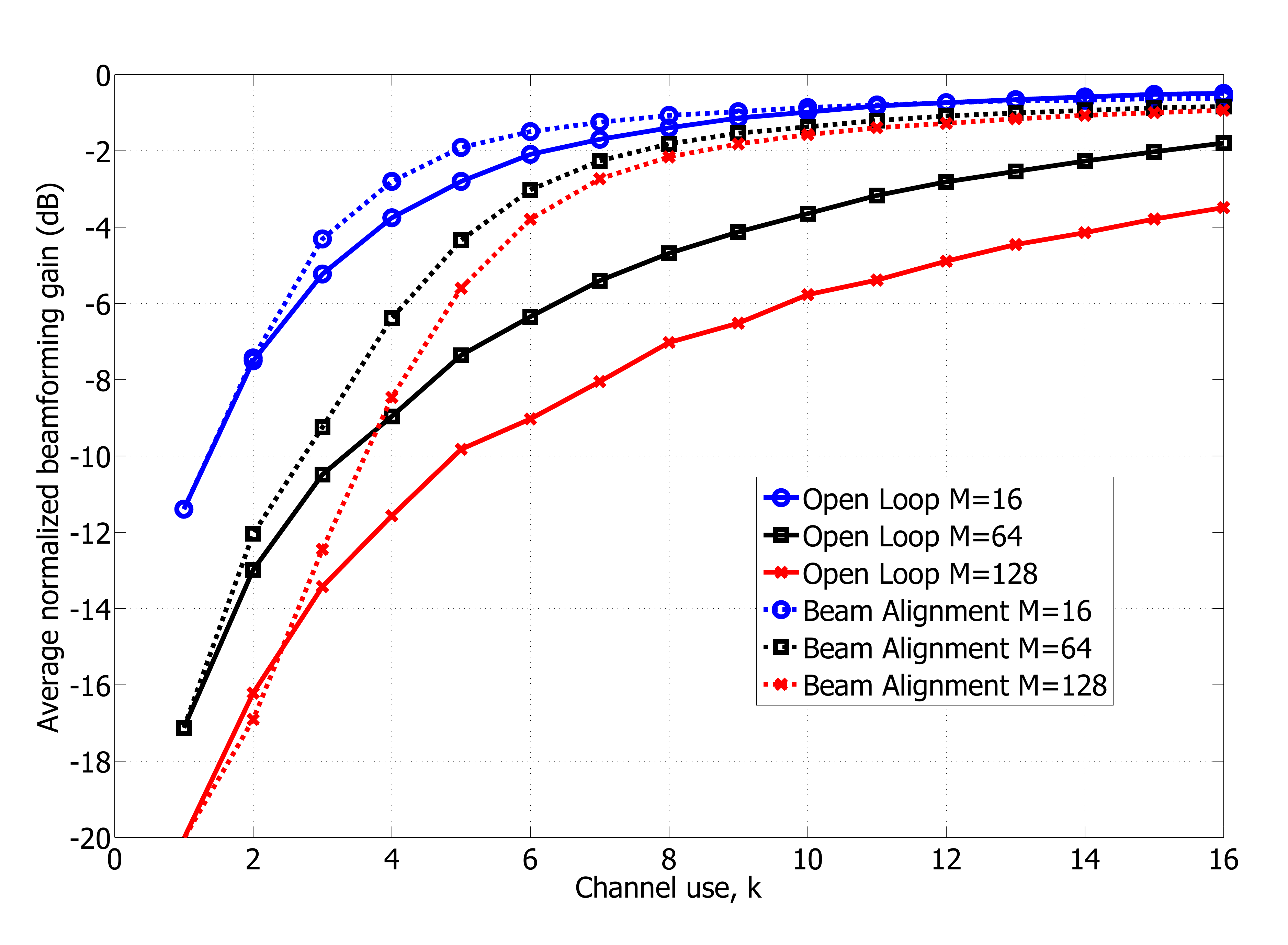}
\caption{Average normalized beamforming gain $\abs{\bh^* \bhhat^{(k)}}^2$ for
a line-of-sight channel.}
\label{fig:los_channel}
\end{figure}

In this section, we present numerical results for shortened training intervals,
e.g. $K < M$.  Consider a line-of-sight channel for a uniform linear array.
Current cellular deployments place basestation antennas on high towers.  In many cases,
the channel between the base station and the user lacks any rich scattering,
making it appropriate to model as a line-of-sight channel. Fig.
\ref{fig:los_channel} shows the average beamforming gain as a function of
channel use for SNR$=0$dB.  We model the channel to be on the array manifold,
$\bh = \ba(\theta)$, where $[\ba(\theta)]_m = \frac{1}{\sqrt{M}} e^{j \pi (m-1)
\sin(\theta)}$. The channel gain term $\alpha$ is randomly chosen as a sum of
$M$ independent zero-mean unit-variance complex Gaussian random variables. To
construct $\cH$, the angle $\theta$ was uniformly quantized over $\theta \in [-\pi,\pi]$.   Results were averaged
over $5000$ iterations and used a channel codebook of size $\abs{\cH} = 2 M$ and a sounding
vector codebook of size $\abs{\cW} = M$. Sounding vectors are chosen by minimizing the misalignment cost function using an approximation
for the updated priors expression. An open-loop scheme is also presented, with
predetermined sounding vectors designed to be the first $K$ columns of the
identity matrix. The open-loop scheme then estimates $\alpha \bh$ using an MMSE
estimator, and quantizes its estimate to the channel codebook $\mathcal{H}$.

The closed-loop
beam alignment algorithm is able to increase the average beamforming gain over
orthogonal training for a small number of channel uses.  At
$k=5$, there is a $0.4$dB gain for $M=16$, a $3$dB gain for
$M=64$, and a $5$dB gain for $M=128$.  The beam alignment scheme gives
considerable improvement for $K/M \ll 1$, and the improvement over open-loop
grows with $M$.

\section{Conclusion}\label{sec:conclusion}
This work developed a closed-loop beam alignment scheme which, through the use
of feedback after each channel use, sequentially designs sounding vectors to
probe the channel in an efficent manner. This scheme provides a practical
channel estimation algorithm for massive MIMO, where large scale arrays make
optimal training impractical. A generalized maximum likelihood detector was
developed to jointly perform the channel estimation and channel quantization. Since beamforming gain
is only concerned with the direction of the channel, the detector replaces the
channel gain by its maximum likelihood estimate. Sounding
vectors are selected to minimize the misalignment cost function, a metric
updated with knowledge of the previous receive samples in a Bayesian
framework. The closed-loop beam alignment scheme shows improved beamforming gain
over conventional orthogonal training signals, especially for the $K<M$.
Additional analysis is required to determine the necessary amount of training as the number of transmit antennas
increases.

\appendix
Consider two  Gaussian random variables, $X = \mu_x + q_x n$ and $Y = \mu_y
+ q_y n$. It should be mentioned that the expression $\text{Pr}(\abs{X}^2 >
\abs{Y}^2)$ in Appendix B of \cite{Proakis} is not applicable when both $X$ and
$Y$ are functions of the same scalar noise term $n \sim \mathcal{CN}(0,1)$, in
which for a scalar noise term $n$, $E[\abs{X-\mu_x}^2]
E[\abs{Y-\mu_y}^2] - \abs{E\left[(X-\mu_x)(Y-\mu_y)^*\right]}^2 = 0$.

The inequality inside the probability expression, $\abs{\mu_x + q_x n}^2 >
	\abs{\mu_y + q_y n}^2$, can be simplified by
completing the square and rearranging terms for three separate cases (which
depend on the magnitudes of $q_x$ and $q_y$),
\begin{IEEEeqnarray*}{rl}
	\IEEEeqnarraymulticol{2}{l}{\text{Pr}\left( \abs{\mu_x + q_x n}^2 >
	\abs{\mu_y + q_y n}^2 \right) = }\\
	\qquad 
	\begin{cases}
		\text{Pr}\left( \abs{Z}^2 < \left\vert \frac{\mu_x q_y - q_x
		\mu_y}{\abs{q_y}^2 - \abs{q_x}^2} \right\vert^2 \right) & \text{ if }
		\abs{q_y}^2 > \abs{q_x}^2\\
		\text{Pr}\left( \abs{Z}^2 > \left\vert \frac{\mu_x q_y - q_x
		\mu_y}{\abs{q_y}^2 - \abs{q_x}^2} \right\vert^2 \right) & \text{ if }
		\abs{q_y}^2 < \abs{q_x}^2\\
		\text{Pr}\left( V < \frac{\abs{\mu_y}^2 - \abs{\mu_x}^2}{2} \right) & \text{
		if } \abs{q_y}^2 = \abs{q_x}^2\\
	\end{cases}
\end{IEEEeqnarray*}
where $Z = n + \frac{\mu_y q_y^* - \mu_x q_x^*}{\abs{q_y}^2 -
	\abs{q_x}^2}$ is a complex Gaussian random variable and
$V = \text{Re}\{\mu_y^* q_y - \mu_x^* q_x\} \text{Re}\{n\} - \text{Im}\{\mu_y^* q_y
- \mu_x^* q_x\} \text{Im}\{n\}$ is a real Gaussian random variable. $\abs{Z}^2$
is a noncentral chi-squared random variable with  $\text{Pr}(\abs{Z}^2 < z) = 1
- Q_1(\abs{\lambda}, \sqrt{z} )$ and $V$ is the sum
of two independent real Gaussian random variables, with $\text{Pr}(V < v) =
\frac{1}{2} \left( 1+ \text{erf}\left( v/\sqrt{2 \abs{\mu_y^* q_y -
\mu_x^* q_x}^2} \right) \right)$. We then conclude the result in
\eqref{eq:P_xy}.

\bibliographystyle{IEEEtran}
\bibliography{IEEEabrv,BeamAlignment_bib}

\end{document}